\documentclass[11pt]{article}

\pdfoutput=1
\usepackage[paperwidth=8.5in, paperheight=11in, portrait, top=1in, bottom=1.in,
 left=1.15in, right=1.15in]{geometry}
\usepackage[utf8]{inputenc}
\usepackage{amssymb,amsthm,amsmath,amsfonts,mathtools,bm}
\usepackage{authblk} 
\usepackage{graphicx,mathrsfs,latexsym,xcolor,subfigure,epsfig}
\usepackage{units}
\usepackage[breaklinks]{hyperref}

    \def\cB{{\cal B}}    
\def\cD{{\cal D}}        \def\cF{{\cal F}}

\def\fc{ \mathfrak{c} }
\def\ff{ \mathfrak{f} }
\def\fs{ \mathfrak{s} }
\def\CC{ \mathbb{C}}

\numberwithin{equation}{section}
\theoremstyle{plain}

\title{\bf Time and band limiting operator and Bethe ansatz}
\renewcommand*{\Affilfont}{\normalsize\small}
\author[1]{Pierre-Antoine Bernard\,}
\author[2]{Nicolas Crampé\,}
\author[3]{Luc Vinet\,\vspace{.5em}}
\affil[1,3]{Centre de Recherches Math\'ematiques, Universit\'e de Montr\'eal,
\newline\vspace{.9em}
P.O. Box 6128, Centre-ville Station, Montr\'eal (Qu\'ebec), H3C 3J7, Canada.}
\affil[2]{Institut Denis-Poisson CNRS/UMR 7013 - Université de Tours - Université
d'Orléans, \newline\vspace{.9em}
Parc de Grandmont, 37200 Tours, France.}
\affil[3]{Insitut de valorisation des donn\'ees (IVADO), Montr\'eal (Qu\'ebec), H2S 3H1, Canada. \newline\vspace{.9em}}

{
 \makeatletter
 \renewcommand\AB@affilsepx{: \protect\Affilfont}
 \makeatother
 \affil[ ]{E-mail addresses}
 \makeatletter
 \renewcommand\AB@affilsepx{, \protect\Affilfont}
 \makeatother
 \affil[1]{pierre-antoine.bernard@umontreal.ca}
 \affil[2]{crampe1977@gmail.com}
 \affil[3]{vinet@crm.umontreal.ca}
}
\date{\today}

\begin{document}
\maketitle

\hrule
\begin{abstract}
The time and band limiting operator is introduced to optimize the reconstruction of a signal from only a partial part of its spectrum.
In the discrete case, this operator commutes with the so-called algebraic Heun operator which appears in the context of the quantum integrable systems.
The construction of both operators and the proof of their commutativity is recalled. A direct connection between their spectra is obtained. 
Then, the Bethe ansatz, a well-known method to diagonalize integrable quantum Hamiltonians, 
is used to diagonalize the Heun operator and to obtain insights on the spectrum of the time and band limiting operator. 

\end{abstract}
\hrule

\section*{Introduction}

The reconstruction of a signal in a finite interval of time when only a part of the Fourier data is available leads to the diagonalization of the so-called time and band limiting operator. 
This paper establishes the link with the time and band limiting operator in the discrete case, the Heun operator and the Bethe ansatz. 

The relation between the 
time and band limiting operator and the Heun operator was first highlighted in the continuous case. Slepian, Landau and Pollack considered the following question: how much can a function limited 
to a band of frequencies $[-\Omega,\Omega]$ be concentrated in the finite interval $[-t,t]$? One might as well consider the problem where the role of time and frequency are exchanged. 
In a seminal series of paper \cite{pswf1,pswf2}, they showed that an answer could be obtained by diagonalizing an integral operator with a $sinc$ kernel, 
\textit{i.e.} the time and band limiting operator. 
They also made the surprising observation that the latter has the property of commuting with a second order linear differential operator and thus 
shares with it a common set of eigenfunctions. These were further recognized to be the prolate spheroidal wave functions. 
This breakthrough greatly simplifies the diagonalization of the integral operator, which is otherwise difficult even numerically. 

These results have now been applied in various settings, from random matrix theory \cite{Cloi1, Meht1} to the study of quantum entanglement in free fermions systems \cite{Eis1}. 
They have also been generalized to treat cases where the functions are defined on finite \cite{Grun1} and infinite discrete sets, on circles and in higher dimensions \cite{pswf3, pswf4, pswf5}. 
Furthermore, other instances in which an integral operator or a full matrix commutes with a simple differential operator or tridiagonal matrix have since been discovered. 
In many cases, including the original $sinc$ kernel and its discrete counterpart, the commuting operator can be identified as an algebraic Heun operator \cite{GVZ, GVZ2}. 
It has been also used for the computation of the entanglement entropy of free Fermions on different chains \cite{CNV,CNV2,BCSV} and on graphs associated to various association schemes
\cite{CGV,BCV,BCV2}. These results provide an algebraic explanation of the existence of a tridiagonal matrix commuting with $Q^\pm$ which has been 
found by direct computations \cite{EP}.

Recently, the Heun operator of Askey--Wilson type has been identified in the transfer matrix associated to the $XXZ$ spin chain with generic boundaries thus allowing to use 
the methods developed in this context to study the Heun operator \cite{BasP}. Although the integrability of the $XXZ$ spin chain with generic boundaries was discover already in 1988 \cite{Skl},
its diagonalization was an open problem for a long time. The eigenvalues in the context of the Bethe ansatz have been given only 25 years later in \cite{Cao1,Cao2,Nepo} and the eigenvectors
were obtained in \cite{BellP} by using a generalization of the Bethe ansatz called the modified algebraic Bethe ansatz.  Let us also mention that the separation of variable method has been
developed in parallel to solve the same problem \cite{Nic}. This connection between integrable systems and Heun operators opens a very interesting path and has already been used in the context of the computation
of the entanglement entropy \cite{BCSV}. In this paper, we show that it allows to get insights into the time and band limiting problem by providing the Bethe ansatz for its associated Heun operator.

The paper is organized as follows. In Section \ref{sec:DTLO}, the discrete Fourier transform is recalled to fix the notations. The time and band limiting problem is settled precisely with the definition
of the time and band limiting operator. The associated Heun operator is then constructed and it is shown that the time and band limiting operator can be expressed as a polynomial of the latter. 
Section \ref{sec:bet} is devoted to the diagonalization of the Heun operator by the Bethe ansatz method. 
Although the results could be deduced from \cite{BasP} as different limits, we give a self-contained presentation for the particular case needed in this paper. 
We do not use the $R$-matrix formalism introduced usually in the algebraic Bethe ansatz but give a straightforward presentation using only the Askey--Wilson algebra. 
We believe that this presentation is simpler to follow for the non-experts of the Bethe ansatz methods.

\section{Discrete time and band limiting operators \label{sec:DTLO}}

\subsection{Fourier transform}

Let $\cF$ denote the vector space over the complex field consisting of all the functions from $\{0,1,2,\dots, 2n-1\}$ to $\CC$. 
Its dimension is $\text{dim}(\cF)=2n$. Only the even case is considered in this paper to simplify its presentation although similar results can be obtained for the odd case.
For two functions $f,g\in \cF$, a scalar product is defined by
\begin{equation}
 \langle f| g \rangle=\sum_{x=0}^{2n-1} f(x)^* g(x)\,.
\end{equation}
The space $\cF$ can be decomposed as a direct sum $\cF=\cF^+   \oplus \cF^-$ where a function $f$ belongs to $\cF^\pm$
if it has the symmetry property: $f(j)=\pm f(2n-j)$ for any $j\in \{0,1,2,\dots, 2n-1\}$ (by convention $f(2n)=f(0)$). The dimensions of these vector 
spaces are $\text{dim}(\cF^+)=n+1$ and $\text{dim}(\cF^-)=n-1$.

For latter convenience, let us introduce the functions $\delta_k$, $\fs_k$, and $\fc_k$ of $\cF$ defined by 
\begin{eqnarray}
 \delta_k(x)=\begin{cases}
              1, &\text{if } k=x \text{ mod } 2n\\
              0,& \text{otherwise}
             \end{cases},
\end{eqnarray}
and
\begin{eqnarray}
 \fs_k(x)=\sin\left(\frac{\pi kx}{2n}\right)\,,\qquad
\fc_k(x)=\cos\left(\frac{\pi kx}{2n}\right)\, .
\end{eqnarray}
Let us notice that $\delta_{2n}=\delta_0$,  $\fs_{4n}=\fs_0$ and $\fc_{4n}=\fc_0$.
We introduce also the following shortened notations:
\begin{equation}
 \fc=\fc_1  \,, \quad  \fs=\fs_1 \,.
\end{equation}

Orthonormal bases of  $\cF^+$ and $\cF^-$ are given by the following sets of vectors, respectively,   
\begin{equation}
  |j,+\rangle= \frac{ \delta_j + \delta_{2n-j}}{\rho(j)\sqrt{2}} , \quad \text{for } 0\leq j \leq n,\quad\text{and}\quad |j,-\rangle=\frac{ \delta_j - \delta_{2n-j}}{\sqrt{2}},\quad   \text{for } 1\leq j\leq n-1\,,
\end{equation}
where the function $\rho(j)=1$ if $1\leq j \leq n-1$ and $\rho(0)=\rho(n)=\sqrt{2}$.
These bases are called position bases. 
There exist other natural bases, called momentum bases.
The orthonormal vectors for the momentum bases of $\cF^+$ and $\cF^-$ are, respectively,  
\begin{equation}
 |\theta_k,+\rangle= \frac{ \fc_{2k}}{\rho(k)\sqrt{n} }\,\quad\text{for } 0\leq k\leq n,\quad\text{and}\quad   |\theta_k,-\rangle= \frac{ \fs_{2k}}{\sqrt{n} },\quad   \text{for } 1 \leq k \leq n-1\,.
\end{equation}
By convention, we set $ |0,-\rangle=|n,-\rangle=|\theta_0,-\rangle=|\theta_n,-\rangle=0$.

The Fourier transform consists in writing a function $f$ of $\cF$ in terms of $|\theta_k,+\rangle$ and $|\theta_k,-\rangle$:
\begin{equation}
 |f \rangle\sum_{k=0}^n \ff^+_k|\theta_k,+\rangle + \sum_{k=1}^{n-1} \ff^-_k  |\theta_k,-\rangle\,,
\end{equation}
with $\ff^\pm_k \in \CC$ the Fourier coefficients.
In particular, the Fourier transforms of the position basis vectors are, for  $0 \leq j,k\leq n$, 
\begin{eqnarray}
| j,+\rangle=\sqrt{\frac{2}{n}} \sum_{k=0}^n  \frac{\fc_{2k}(j) }{\rho(k)\rho(j)}  |\theta_k,+\rangle\,, \qquad  | j,-\rangle= \sqrt{\frac{2}{n}} \sum_{k=1}^{n-1}\fs_{2k}(j)  |\theta_k,-\rangle.
\end{eqnarray}

\subsection{The discrete time and band limiting problem and $Q^\pm$}

As indicated in the introduction, 
an important challenge in signal processing is to reconstruct a function restricted to a finite interval from an incomplete knowledge of its Fourier decomposition. For the finite discrete case introduced in \cite{Grun1}, one wants to recover $f$ given that:
\begin{itemize}
\item $\ff_k^{\pm}$ is only known for $ k \in \{0,1,\dots, K\} \subset \{0,1,\dots, n\}$;
\item $f(x) = 0$ outside the interval $x \in \{-L, \dots, L-1, L\} $ with $0\leq L \leq n$ (where $x$ must be understood modulo $2n$).
\end{itemize}
In terms of the projectors
\begin{equation}
 \pi_1^\pm =\sum_{j=0}^L |j,\pm\rangle\langle j,\pm|\,, \qquad  \pi_2^\pm =\sum_{k=0}^K |\theta_k,\pm\rangle\langle \theta_k,\pm | \,,
\end{equation}
it amounts to determining $ f = (\pi_1^+ + \pi_1^-) f$ from the knowledge of
$(\pi_2^+ \pi_1^+ + \pi_2^- \pi_1^- )f$. This is referred to as the discrete time and band limiting problem. To assert if this is possible, one has to consider the singular value decomposition of $E_\pm = \pi^\pm_2 \pi^\pm_1$. This decomposition can be deduced from diagonalization of the (discrete) time and band limiting operators,
\begin{eqnarray}
 Q^\pm = E_\pm^* E_\pm =\pi^\pm_1 \pi^\pm_2 \pi^\pm_1.
 \label{TBO}
\end{eqnarray}
In particular, if $Q^\pm$ has a zero eigenvalue associated to an eigenvector in $\pi_1^\pm \mathcal{F}^\pm$, 
then the intersection of $\pi_1^\pm \mathcal{F}^\pm$ with the kernel of $E_\pm$ is non-empty and recovering $f$ exactly is impossible. 
The presence of eigenvalues near zero also indicates a risk for numerical instability. 

It is interesting to note that the diagonalization of the time and band limiting operators $Q^\pm$ is motivated by other questions. The eigenvector associated to the largest eigenvalue of $Q^+$ (resp. $Q^-$) gives the symmetric function $g_+ = \pi_1^+ g$ (resp. antisymmetric function $g_- = \pi_1^- g$) restricted to the interval $x \in \{-L, \dots, L-1, L\}$ which is best contained in the band of frequencies $\{0, 1, \dots, K\}$, \textit{i.e.} 
\begin{align}
\frac{\| \pi_2^\pm g_\pm \|}{\| g_\pm \|} = \text{max}_{f \in \pi_1^\pm \mathcal{F}_\pm} \left(\frac{\| \pi_2^\pm f \|}{\| f \|}\right).
\end{align}

The matrices $Q^\pm$ also arise as the chopped correlation matrices for systems of free fermions on $2n$-gon in their ground state. 
Their spectrum therefore contains the necessary information to compute von Neumann entanglement entropies. Indeed, the methods used in \cite{CGV,BCV,BCV2}
associated to different graphs may be used in the context of the $2n$-gon.

Next, we derive by algebraic means the tridiagonal matrices commuting with $Q_\pm$. These were also obtained in \cite{Grun1}.

\subsection{Associated Heun operators}

\paragraph{Leonard pair.}
Let us introduce the operators $A_-$ and $A^*_-$ on $\cF^-$ by
\begin{equation} \label{Aa}
A_-|j,-\rangle=|j-1,-\rangle+|j+1,-\rangle \,, \quad\text{and} \quad A^*_-|j,-\rangle=2\fc (2j)|j,-\rangle\,,
\end{equation}
for $j=1,\dots,n-1$. We recall that $|0,-\rangle=|n,-\rangle=0$.
By direct computation, one can show that $|\theta_k,-\rangle$ are the eigenvectors of $A_-$ and that the action of  $A^*_-$ is tridiagonal:
\begin{equation}\label{Aa2}
A^*_-|\theta_k,-\rangle=|\theta_{k-1},-\rangle+|\theta_{k+1},-\rangle \,, \quad\text{and} \quad A_-|\theta_{k},-\rangle=2\fc (2k)|k,-\rangle\,.
\end{equation}

Let us now define the operators $A_+$ and $A^*_+$ on $\cF^+$ by
\begin{equation} \label{As}
A_+|j,+\rangle=\rho(j-1)\rho(j)|j-1,+\rangle+\rho(j)\rho(j+1)|j+1,+\rangle \,, \quad\text{and} \quad A^*_+|j,+\rangle=2\fc (2j)|j,+\rangle\,,
\end{equation}
for $j=0,\dots,n$ (by convention $\rho(-1)=\rho(n+1)=0$).
By direct computation again, one finds that $|\theta_k,+\rangle$ are the eigenvalues of $A_+$ and that the action of  $A^*_+$ is tridiagonal:
\begin{equation}\label{As2}
A^*_+|\theta_k,+\rangle=\rho(k-1)\rho(k)|\theta_{k-1},+\rangle+\rho(k)\rho(k+1)|\theta_{k+1},+\rangle \,, \quad\text{and} \quad A_+|\theta_{k},+\rangle=2\fc (2k)|k,+\rangle\,.
\end{equation}

The couples of operators $(A_+,A_+^*)$ or $(A_-,A_-^*)$ satisfying the above properties are usually called Leonard pair \cite{Ter,TV}.

\paragraph{Heun operators.} For a given Leonard pair $(A,A^*)$, the associated algebraic Heun operator is the most general bilinear combination of $A$ and $A^*$ \cite{NT,GVZ}
\begin{equation} \label{eq:Heun}
 T=r_1\{A,A^*\} +r_2[A,A^*] +r_3 A^* +r_4 A +r_5\ ,
\end{equation}
where $r_i$ are new parameters. It has the property of acting tridiagonally on the eigenbasis of both $A$ and $A^*$.
We are interested in the particular Heun operators given by
\begin{equation} \label{eq:Heunp}
 T^\pm=\frac{1}{4\fc(1)}\{A_\pm,A_\pm^*\} - \fc(2K+1) A_\pm^* -\fc(2L+1) A_\pm\,.
\end{equation}
The action of $T^\pm$ on the position and momentum bases can be deduced from equation \eqref{Aa}-\eqref{As2}. On the position basis, this action reads explicitly as follows:
\begin{align}
T^\pm |j,\pm \rangle = a_{j,\pm}|j-1,\pm \rangle + b_{j,\pm}|j,\pm \rangle + c_{j,\pm}|j+1,\pm \rangle,
\label{AcHeun}
\end{align}
where
\begin{align}
a_{j+1,-} = c_{j, -} = \fc(2j + 1) - \fc(2L+1), \quad  b_{j, \pm} = -2\fc(2K + 1) \fc(2j),
\end{align}
\begin{align}
a_{j+1,+} = c_{j, +} = \rho(j)\rho(j+1)(\fc(2j + 1) - \fc(2L+1) ).
\end{align}
On the momentum basis, it is:
\begin{align}
T^\pm |\theta_k,\pm \rangle = \overline{a}_{k,\pm}|\theta_{k-1},\pm \rangle + \overline{b}_{k,\pm}|\theta_k,\pm \rangle + \overline{c}_{k,\pm}|\theta_{k+1},\pm \rangle,
\end{align}
where
\begin{align}
\overline{a}_{k+1,-} = \overline{c}_{k, -} = \fc(2k + 1) - \fc(2K+1), \quad  \overline{b}_{k, \pm} = -2\fc(2L + 1) \fc(2k),
\end{align}
\begin{align}
\overline{a}_{k+1,+} = \overline{c}_{k, +} = \rho(k)\rho(k+1)(\fc(2k + 1) - \fc(2K+1) ).
\end{align}
Note that $[T^\pm, \pi_1^\pm]=0$ also implies that we can choose the basis $|t_{\ell, \pm}\rangle$ such that the first $L + 1$ vectors are in the subspace onto which $\pi_1^\pm$ projects, \textit{i.e.}
\begin{align}
\pi_1^\pm | t_{\ell,\pm} \rangle =
  \left\{
    \begin{array}{ll}
    	 | t_{\ell,\pm} \rangle  & \mbox{if } \ell \leq L,  \\
    	0 & \mbox{if } \ell > L.
    \end{array}
\right. 
\end{align}

From these explicit forms of $T^\pm$, it is easy to show that they satisfy $[T^\pm, \pi_1^\pm]=0$ and $[T^\pm, \pi_2^\pm]=0$. Therefore, $T^\pm$ 
commute with the time and band limiting operators $Q^\pm$.
They can be diagonalized in the same basis $\{| t_{\ell, \pm} \rangle \}$, with ${\ell \in \{0,1,\dots,\text{dim}(\mathcal{F}^\pm)-1\}}$:
\begin{align}
T^\pm | t_{\ell,\pm} \rangle = t_{\ell,\pm} | t_{\ell, \pm} \rangle, \quad \quad Q^\pm | t_{\ell,\pm} \rangle = q_{\ell,\pm} | t_{\ell, \pm} \rangle.
\label{diag}
\end{align}

\subsection{The time-band limiting operators as polynomials of the Heun operators}

We are now interested in the relation between $t_{\ell, \pm}$ and $q_{\ell, \pm}$. In other words, we are looking 
to express the time and band limiting operators $Q^\pm$ as polynomials $P_\pm$ of $\pi_1^\pm T^\pm$, 
\textit{i.e.}
\begin{align}
Q^\pm = \pi_1^\pm P_\pm(T^\pm).
\end{align}
In particular, we look for polynomials $P_\pm$ such that 
 \begin{align}
 P_+(t_{\ell, +}) = q_{\ell, +}, \quad \forall \ell \in \{0, 1, \dots, L\}
 \end{align}
 and
  \begin{align}
 P_-(t_{\ell, -}) = q_{\ell, -}, \quad \forall \ell \in \{0, 1, \dots, L-1\}.
 \end{align}
This would allow to obtain the spectrum of $Q^\pm$ from the eigenvalues $t_{\ell, \pm}$ of $T^\pm$, $\ell \in \{0,1, \dots, L\}$.

 This type of relation between the spectra of the Heun and the time and band limiting operators has been discovered previously in different contexts \cite{BCSV,BCV3}. We follow the approach used in \cite{BCV3}, where the continuous time and band limiting operator is expressed as function of the confluent Heun operator,  and consider the following equation:
\begin{align}
\left(T^\pm \otimes I  - I \otimes T^\pm \right) |\psi \rangle = 0,
\label{HeunHeun}
\end{align}
where $|\psi \rangle \in \mathbb{C}^{\text{dim}(\mathcal{F}^\pm)} \otimes \mathbb{C}^{\text{dim}(\mathcal{F}^\pm)} $. We also define $E_{j, \pm}$ as the following matrices
\begin{align}
E_{j, +} = |0,+ \rangle \langle j,+ | \otimes I, \quad E_{j, -} = |1,- \rangle \langle j+1,- | \otimes I,
\end{align}
where $I$ is the identity matrix. For any solution $|\psi \rangle$ of \eqref{HeunHeun}, we note that
\begin{equation}
	\begin{split}
(I \otimes T^\pm) E_{j, \pm} |\psi \rangle 
 =  E_{j, \pm} (I \otimes T^\pm) |\psi \rangle = E_{j, \pm} (T^\pm \otimes I) |\psi \rangle.
	\end{split}
\end{equation}
 Then, we can use \eqref{AcHeun} and the Schmidt decomposition
\begin{align}
|\psi \rangle = \sum_{j = 0}^n | j, \pm  \rangle \otimes | \chi_{j,\pm} \rangle,
\end{align}
to show that
\begin{equation}
	\begin{split}
(I \otimes T^\pm) E_{j, +} |\psi \rangle 
 & = E_{j, +} \sum_{j' = 0}^n \big(a_{j', +} |{j'-1,+} \rangle +b_{j', +} |{j',+} \rangle + c_{j', +} |{j'+1,+} \rangle\big) \otimes | \chi_{j',+} \rangle \\
 & =  a_{j+1, +} E_{j+1, +} |\psi \rangle +  b_{j, +}E_{j, +} |\psi \rangle +  c_{j-1, +}E_{j-1, +} |\psi \rangle
	\end{split}
\end{equation}
and
\begin{equation}
	\begin{split}
(I \otimes T^\pm) E_{j, -} |\psi \rangle 
 & =  a_{j+2, -} E_{j+1, -} |\psi \rangle +  b_{j+1, -}E_{j, -} |\psi \rangle +  c_{j, -}E_{j-1, -} |\psi \rangle.
	\end{split}
\end{equation}
For $j \in \{0, 1, \dots, L - 1\}$, we have $a_{j+1,\pm} \neq 0$ and the previous expression yields a three-term recurrence relation for the action of the matrix $E_{j,\pm}$ on $|\psi\rangle$. Thus, we can deduce that 
\begin{align}
 E_{j,\pm} |\psi \rangle =  R_{j,\pm}(I \otimes T^\pm)E_{0,\pm} |\psi \rangle,
\label{Poly}
\end{align}
where $R_{j,+}$ is a polynomial of order $j \in \{0, 1, \dots, L \} $ which verifies
\begin{align}
x R_{j,+}(x)  = a_{j+1, +}R_{j+1,+}(x) +  b_{j, +}R_{j,+}(x) +  c_{j-1, +}R_{j-1,+}(x), \quad R_{0,-}(x) =  1
\end{align}
and  $R_{j,-}$ is a polynomial of order $j \in \{0, 1, \dots, L-1 \} $ 
which verifies
\begin{align}
x R_{j,-}(x)  = a_{j+2, -}R_{j+1,-}(x) +  b_{j+1, -}R_{j,-}(x) +  c_{j, -}R_{j-1,-}(x), \quad R_{0,-}(x) =  1
\end{align}
It is interesting to note that equation \eqref{Poly} can also be rewritten as
\begin{align}
R_{j,+}(T^+) | \chi_{0,+}\rangle = | \chi_{j,+}\rangle, \quad R_{j,-}(T^-) | \chi_{1,-}\rangle = | \chi_{j+1,-}\rangle.
\end{align}
Therefore, the matrices $R_{j,\pm}(T^\pm)$ have an action that can be interpreted as a translation and play a role similar to the operators $U(\xi ; T)$ introduced in \cite{BCV3}. Next, we can use the polynomials $R_{j,\pm}$ to construct $P_\pm$. Note that
\begin{align}
| \psi_\ell \rangle = {| t_{\ell, \pm} \rangle \otimes | t_{\ell, \pm} \rangle}, \quad \ell \in \{0,1,\dots,L\},
\end{align}
verifies equation \eqref{HeunHeun}. Thus, equation \eqref{Poly} can be applied to this vector to obtain
\begin{align}
\langle 0, + | t_{\ell, +} \rangle R_{j,+}(I \otimes T^+) | 0,+ \rangle \otimes   | t_{\ell, +} \rangle = {\langle j, + | t_{\ell, +} \rangle} | 0,+ \rangle \otimes | t_{\ell, +} \rangle,
\end{align}
and
\begin{align}
\langle 1, - | t_{\ell, -} \rangle R_{j,-}(I \otimes T^-) | 1,- \rangle \otimes   | t_{\ell, -} \rangle = {\langle j+1, - | t_{\ell, -} \rangle} | 1,- \rangle \otimes | t_{\ell, -} \rangle.
\end{align}
So, one concludes that
\begin{align}
R_{j,+}(t_{\ell, +} ) = \frac{\langle j, + | t_{\ell, +} \rangle}{\langle 0, + | t_{\ell, +} \rangle}, \quad R_{j,-}(t_{\ell, -} ) = \frac{\langle j+1, - | t_{\ell, -} \rangle}{\langle 1, - | t_{\ell, -} \rangle}.
\end{align}
Since $|t_{\ell, \pm}\rangle$ is an eigenvector of $Q^\pm$, one can check that the polynomials
\begin{align}
P_+ = \sum_{j = 0}^{L} \langle 0, + | Q^+ | j,+ \rangle R_{j,+}, \quad P_- = \sum_{j = 0}^{L-1} \langle 1, - | Q^- | j+1,- \rangle R_{j,-}
\end{align}
verify respectively
\begin{align}
P_+(t_{\ell, +}) = \sum_{j = 0}^{L} \langle 0, + | Q^+ | j,+ \rangle \frac{\langle j, + | t_{\ell, +} \rangle}{\langle 0, + | t_{\ell, +} \rangle} = \frac{\langle 0, + | Q^+ | t_{\ell, +} \rangle}{\langle 0, + | t_{\ell, +} \rangle} = q_{\ell, +},
\end{align}
for $\ell \in \{0,1,\dots,L\}$, and
\begin{align}
P_-(t_{\ell, -}) = \sum_{j = 0}^{L-1} \langle 1, - | Q^- | j+1,- \rangle \frac{\langle j+1, - | t_{\ell, -} \rangle}{\langle 1, - | t_{\ell, -} \rangle} = \frac{\langle 1, - | Q^- | t_{\ell, -} \rangle}{\langle 1, - | t_{\ell, -} \rangle} = q_{\ell, -},
\end{align}
for $\ell \in \{0,1,\dots,L-1\}$. It also follows that
\begin{align}
Q^\pm = \pi_1^\pm P_\pm(T^\pm).
\end{align}

\section{Bethe ansatz \label{sec:bet}}

\paragraph{Askey--Wilson algebra.}
The previous pairs of operators $(A_+,A_+^*)$ or $(A_-,A_-^*)$ satisfy the following relations
\begin{eqnarray} \label{eq:terA}
 &&A_\pm^2A_\pm^*-2\fc(2) A_\pm A_\pm^*A_\pm +A_\pm^*A_\pm^2=4\fs(2)^2 A_\pm^*\,, \\ 
 &&(A_\pm^*)^2A_\pm-2\fc(2) A_\pm A_\pm^*A_\pm +A_\pm(A_\pm^*)^2=4\fs(2)^2 A_\pm\,. \label{eq:terA2}
\end{eqnarray}
These relations define an algebra which is a particular case of the Askey--Wilson algebra appearing in the context of the eponymous polynomials \cite{Zhe1} or in the Racah problem 
for $U_q(sl_2)$ \cite{GZ} (see \textit{e.g.} \cite{CFGPRV} for a review).
In the context of association schemes, this algebra corresponds to the irreducible decomposition of the Terwilliger algebra associated to the $2n$-gon.

It has been found to also appear naturally in the framework of the integrable systems \cite{bas1,bas2,bas3} as a realization of the reflection equation \cite{Skl}.
It has further been shown that it is possible to diagonalize the Heun operators associated to the
general Askey--Wilson algebra \cite{BasP} and to the particular case of the Lie algebra $sl_2$ \cite{BCSV}.
These diagonalizations rely on the modified Bethe ansatz which has been developed to deal with quantum integrable models without $U(1)$ symmetry 
\cite{BC,BellP}. 

In the following, we show how the Bethe ansatz can be used to diagonalize the Heun operators $T^\pm$ associated to the algebra \eqref{eq:terA}-\eqref{eq:terA2}. 
Usually, the $R$-matrix formalism is much called upon. 
However, we chose to employ here only the commutation relations \eqref{eq:terA}-\eqref{eq:terA2} even if the definitions below are inspired  by the $R$-matrix formalism.

\paragraph{Dynamical operators.} 
Let us introduce the following dynamical operators:
\begin{eqnarray}
 &&\hspace{-0.8cm}\cD_\pm(u,m)=\frac{1}{\fs(2u)\fs(2m)}\left(  \fc(2m+1) A_\pm - \fc(2u-2m) A_\pm^* -\frac{\{A_\pm ,A_\pm^*\}}{4\fc(1)} +2\fc(2u) \right), \label{eq:defD}\\
&& \hspace{-0.8cm}\cB_\pm(u,m)=\frac{1}{\fs(2m)}\left( \fc(1) A_\pm - \fc(2u) A_\pm^* +\frac{\fs(2m)[A_\pm,A_\pm^*]}{4\fs(1)}-\frac{\fc(2m)\{A_\pm,A_\pm^*\}}{4\fc(1)} +2\fc(2u)\fc(2m) \right),\nonumber 
 \end{eqnarray}
 where $m$ is an integer and $u\in \CC$.
 The idea of dynamical operators has been introduced in \cite{Bax, FT} to study the XYZ model; it was then adapted to analyze the $XXZ$ spin chain 
 with boundary fields that are not parallel \cite{Cao}.
 
These operators have nice properties. From the commutation relations \eqref{eq:terA}-\eqref{eq:terA2}, one can show that the dynamical operators satisfy the relations:
 \begin{eqnarray}
  \cB_\pm(u,m) \cB_\pm(v,m-1) &=&  \cB_\pm(v,m) \cB_\pm(u,m-1)\,,\label{eq:relBB}\\
 \cD_\pm(u,m) \cB_\pm(v,m) &=& f(u,v)\cB_\pm(v,m) \cD_\pm(u,m-1)\label{eq:relDB}\\
 && +  \cB_\pm(u,m) \Big( g(u,v,m) \cD_\pm(v,m-1) 
 +g(u,-v,m) \cD_\pm(-v,m-1)\Big)\,,\nonumber
 \end{eqnarray}
where
\begin{eqnarray}
 f(u,v)=\frac{\fs(u+v-1)\fs(u-v-1)}{\fs(u+v)\fs(u-v)} \,,\quad g(u,v,m)=\frac{\fs(1)\fs(2v-1)\fs(2m+v-u)}{\fs(2m)\fs(2u)\fs(u-v)}\,.
\end{eqnarray}

 The Heun operators $T^\pm$ can be expressed in terms of the the dynamical operator $\cD_\pm(u,m)$ for two different values of the parameter $m=L$ and $m=-L-1$.
One gets correspondingly the following explicit expressions:
 \begin{eqnarray}
  T^\pm&=&2\fc(2u)+\Delta(u)  \cD_\pm( u,L)+\Delta(-u)  \cD_\pm(-u,L) \,,\label{eq:T1}\\
  T^\pm&=&2\fc(2u)+\Delta(1-u)\cD_\pm(u,-L-1)+\Delta(1+u)\cD_\pm(-u,-L-1) \,,\label{eq:T2}
 \end{eqnarray}
 where 
 \begin{equation}
  \Delta(u)= \fc\big( u+L-K-\frac{1}{2}\big) \fc\big(u+L+K+\frac{1}{2}\big)\,.
 \end{equation}

 Finally, these dynamical operators have nice action on the vectors $|0,+\rangle$ or $|1,-\rangle$. Indeed, using  \eqref{Aa} or  \eqref{As}, one obtains
 \begin{eqnarray}
  \cD_+(u,m)|0,+\rangle &=&-2|0,+\rangle -\frac{1}{\fs(2u)} \cB_+(u,m)|0,+\rangle \,,\label{eq:DV1}\\
  \cD_-(u,m)|1,-\rangle &=& 2\frac{\fs(2-2u)}{\fs(2u)}|1,-\rangle -\frac{\fs(m-1)}{\fs(m+1)\fs(2u)} \cB_-(u,m)|1,-\rangle \,.\label{eq:DV2}
 \end{eqnarray}

 \paragraph{Bethe ansatz for $T^-$.} Let us consider the following vectors, called Bethe states,
 \begin{equation}\label{eq:Vmx}
  V^-(\overline x)= \cB_-(x_1,L)\cB_-(x_2,L-1) \dots \cB_-(x_{L-2},3) \cB_-(x_{L-1},2)  |1,-\rangle
 \end{equation}
 where $\overline x=\{x_1,\dots x_{L-1}\}$ are some parameters, called Bethe roots.
 Using relation \eqref{eq:relBB}, it is easy to show that  $V^-(\overline x)$ does not depend on the order of the parameters $x_i$.
 Noticing that $\cB_-(x,m)$ has a tridiagonal action on the vectors $\{|j,-\rangle \}$, we deduce that 
 \begin{equation}
  V^-(\overline x)  \in \text{span}_\CC ( \{ |1,-\rangle, |2,-\rangle , \dots |L,-\rangle \} )\,.
 \end{equation}
which corresponds to $\pi_1^-\cF^-$. For different choices of the set $\overline x$, one can expect that the associated vector $ V^-(\overline x)$ are independent.
Therefore, a basis of $\pi_1^-\cF^-$ can be obtained from $L$ independent vectors $V^-(\overline x)$. 
The next step consists in finding the set $\overline x$ such that $V^-(\overline x)$ is an eigenvector $|t_{\ell, -}\rangle$ of $T^-$ (which, we recall, leaves $\pi_1^-\cF^-$ stable).
 
 The results of the previous paragraph permit to compute the action of $\cD_-(u,L)$ on $V^-(\overline x)$. Namely, after algebraic manipulations typical of the algebraic Bethe ansatz approach, 
 one gets
\begin{eqnarray}\label{eq:DBE1}
\cD_-(u,L) V^-(\overline x)& =& 2\frac{\fs(2-2u)}{\fs(2u)} \prod_{i=1}^{L-1} f(u,x_i) V^-(\overline x)\\
                              && + 2\sum_{j=1}^{L-1}\sum_{\epsilon=\pm 1} \frac{\fs(2- 2\epsilon x_j)}{\fs( 2\epsilon x_j)} g(u,\epsilon x_j,L)
                                        \prod_{\genfrac{}{}{0pt}{}{i=1}{i\neq j}}^{L-1} f(\epsilon x_j,x_i) 
                                V^-(u,\overline x_{\neq j})\,, \nonumber
 \end{eqnarray}
where $V^-(u,\overline x_{\neq j})$ stands for $V^-(\overline x)$ with $x_j$ replaced by $u$.
Let us remark that we used relation \eqref{eq:DV2} with the particular value $m=1$ for which the second term in the r.h.s. vanishes.

Then, using expression \eqref{eq:T1} of $T^-$, one deduces its action on $V^-(\overline x)$:
\begin{eqnarray}
T^- V^-(\overline x)& =&\left(2\fc(2u) +2\sum_{\epsilon=\pm}\Delta(\epsilon u) \frac{\fs(2-2\epsilon u)}{\fs(2\epsilon u)} \prod_{i=1}^{L-1} f(\epsilon u,x_i) \right) V^-(\overline x)\\
                              && +2\fs(1) \sum_{j=1}^{L-1} \sum_{\epsilon=\pm 1} \frac{\Delta(\epsilon x_j)\fs(2-2\epsilon x_j)\fs(2\epsilon x_j-1)}{\fs(u-x_j)\fs(u+x_j)\fs(2\epsilon x_j)}
                                        \prod_{\genfrac{}{}{0pt}{}{i=1}{i\neq j}}^{L-1} f(\epsilon x_j,x_i) 
                                V^-(u,\overline x_{\neq j}) \,. \nonumber
 \end{eqnarray}
Finally, asking that the different coefficients in the sum over $j$ in the second row of the previous equation vanish, one gets the following equations for the Bethe roots $\overline x$ , for $j=1,\dots,L-1$,
\begin{eqnarray}
 \frac{\Delta(x_j)\fs(2-2x_j)\fs(1-2x_j)}{\Delta(-x_j)\fs(2+2x_j)\fs(1+2x_j)}=\prod_{\genfrac{}{}{0pt}{}{i=1}{i\neq j}}^{L-1} \frac{\fs(x_j-x_i+1)\fs(x_j+x_i+1)}{\fs(x_j-x_i-1)\fs(x_j+x_i-1)} \,.
\end{eqnarray}
These equations are called Bethe equations.
For each solution of the Bethe equations, the associated vector $V^-(\overline x)$ become an eigenvector $|t_{\ell, -}\rangle$ of $T^-$ with the eigenvalue 
\begin{equation}\label{eq:taum}
 t_{\ell, -}(u)=2\fc(2u) + 2\Delta(u) \frac{\fs(2- 2 u)}{\fs(2u)} \prod_{i=1}^{L-1} f(u,x_i)-2\Delta(- u) \frac{\fs(2+2 u)}{\fs(2 u)} \prod_{i=1}^{L-1} f(- u,x_i)\,.
\end{equation}
At first glance, it seems that this eigenvalue depends on the parameter $u$. However, writing this eigenvalue in terms of $U=\text{exp}\left(\frac{i\pi u}{2n}\right)$, we see that $\tau^-(U)$
is a rational function w.r.t. $U^2$ and $1/U^2$. We can show easily that all the residues at the apparent poles of this rational function vanish using the Bethe equations satisfied by $\overline x$.
It proves that it is a polynomial w.r.t. $U^2$ and $1/U^2$. Then, its asymptotic behavior for $U \to \infty$ is
\begin{equation}
 t_{\ell, -}(U) \underset{U\to \infty}{\sim} U^2 - U^2 \exp(i\pi L/n ) \exp(-i\pi/n)  \prod_{i=1}^{L-1} \exp(-i\pi/n)  
\end{equation}
which actually goes to zero. A similar result holds for the asymptotic limit at 0 which proves that $\tau^-(u)$ does not depend on $u$, as expected, and is the eigenvalue of $T^-$.\\



\paragraph{Second Bethe ansatz for $T^-$.} In the previous paragraph, relation \eqref{eq:T1} between $T^-$ and $\cD_-(u,L)$ has been used. In this paragraph, we want to use relation 
\eqref{eq:T2} between $T^-$ and $\cD_-(u,-L-1)$. The Bethe states are given by
 \begin{equation}
  W^-(\overline y)= \cB_-(y_1,-L-1)\cB_-(y_2,-L-2) \dots \cB_-(y_{L-2},-2L+2) \cB_-(y_{L-1},-2L+1)  |1,-\rangle\,,
 \end{equation}
 where $\overline y=\{y_1,\dots,y_{L-1}\}$ is the set of Bethe roots in this case.
 The action of $\cD_-(u,-L-1)$ on  $W^-(\overline y)$ is computed as before:
 \begin{eqnarray}
\cD_-(u,-L-1) W^-(\overline y) &=& 2\frac{\fs(2-2u)}{\fs(2u)} \prod_{i=1}^{L-1} f(u,y_i) W^-(\overline y) \\
                             && + 2\sum_{j=1}^{L-1}\sum_{\epsilon=\pm 1}  \frac{\fs(2- 2\epsilon y_j)}{\fs( 2\epsilon y_j)}  g(u,\epsilon y_j,-L-1)
                                    \prod_{\genfrac{}{}{0pt}{}{i=1}{i\neq j}}^{L-1} f(\epsilon y_j,y_i)       W^-(u,\overline y_{\neq j}) \nonumber \\
                              &&     -\frac{\fs(2L+1)\fs(4L)}{\fs(2L-1)\fs(2L+2)\fs(2u)}W^-( \{y_1,\dots,y_{L-1},u\}  ) \,, \nonumber 
                                      \end{eqnarray}
 where 
 \begin{equation}
  W^-( \{y_1,\dots,y_{L-1},u\})=\cB_-(y_1,-L-1)\dots \cB_-(y_{L-1},-2L+1)\cB_-(u,-2L)  |1,-\rangle \,.
 \end{equation}
This relation is more complicated than the one of the previous case \eqref{eq:DBE1} since the action \eqref{eq:DV2} of $\cD_-(u,-2L)$ on $|1,-\rangle$ has now two terms.
It is necessary to compute $W^-( \{y_1,\dots,y_{L-1},u\})$ which is the crucial step of the modified algebraic Bethe ansatz. 
A formula has been conjectured in \cite{BC} and 
proven in different cases in \cite{Cra1,ABGP,Cra2} to explain the inhomogeneous terms appearing in the Bethe equations in \cite{Cao1,Cao2,Nepo}.
One remarks that $\langle L+1,-|\cB_-(x,-L-1)|L,-\rangle=0$. Therefore $W^-( \{y_1,\dots,y_{L-1},u\})$ are still in $\pi_1^-\cF^-$ and can be expressed 
in terms of the $L$ vectors $W^-(\overline y)$ and $ W^-(u,\overline y_{\neq j})$. The unexpected fact in the context of the modified algebraic Bethe ansatz 
is that this expression is simple.
In the case treated here, one gets 
\begin{eqnarray}
&&\hspace{-1.6cm}  W^-( \{y_1,\dots,y_L\} )  \nonumber\\
&&\hspace{-1.6cm}=\frac{2\fs(2L-1)\fs(2L+1)}{\fs(4L)}\sum_{j=1}^{L}\frac{\fs(2y_j(L+1))}{\fs(2y_j)} \prod_{\genfrac{}{}{0pt}{}{i=1}{i\neq j}}^{L}
\frac{1}{4\fs(y_i+y_j)\fs(y_i-y_j)}W^-( \{y_1,\dots,y_L\}_{\neq j}  )  \,.
\end{eqnarray}
In the above formula we replace $u$ by $y_L$ to shorten it.
Then, the computations are similar to the ones of the previous paragraph and the eigenvalues of $T^-$ are
\begin{eqnarray}
 t_{\ell, -}(u)&=&2\fc(2u) + 2\Delta(1-u) \frac{\fs(2- 2 u)}{\fs(2u)} \prod_{i=1}^{L-1} f(u,y_i)-2\Delta(1+ u) \frac{\fs(2+2 u)}{\fs(2 u)} \prod_{i=1}^{L-1} f(- u,y_i)\nonumber\\
 &&-2\frac{\fs(2L+1)^2\fs(2u(L+1))}{\fs(2u)}\prod_{i=1}^{L-1} \frac{1}{4\fs(y_i+u)\fs(y_i-u)} \,, \label{eq:t2}
\end{eqnarray}
with the following Bethe equations
\begin{eqnarray}
  \frac{\Delta(1-y_j)\fs(2-2y_j)\fs(1-2y_j)}{\Delta(1+y_j)\fs(2+2y_j)\fs(1+2y_j)}&=&\prod_{\genfrac{}{}{0pt}{}{i=1}{i\neq j}}^{L-1} \frac{\fs(y_j-y_i+1)\fs(y_j+y_i+1)}{\fs(y_j-y_i-1)\fs(y_j+y_i-1)} \label{eq:be2}\\
  &&\hspace{-3cm}-\frac{\fs(2L+1)^2\fs(2y_j(L+1))\fs(2y_j-1)}{\fs(2y_j+1)\fs(2+2y_j)\Delta(1+y_j)} \prod_{i=1}^{L-1} \frac{1}{4\fs(y_i-y_j+1)\fs(y_i+y_j-1)}  \,.\nonumber
\end{eqnarray}
The second line in the expression \eqref{eq:t2} of the eigenvalues $t_{\ell, -}(u)$ and of the Bethe equations \eqref{eq:be2} are referred as the inhomogeneous terms.

\paragraph{Bethe ansatz for $T^+$.}
In this case, if we consider the expression \eqref{eq:T1} connecting $T^+$ and $\cD_+(u,L)$, the corresponding Bethe state is 
 \begin{equation}
  \cB_+(x_0,L)\cB_+(x_1,L-1) \dots \cB_+(x_{L-2},2) \cB_+(x_{L-1},1)  |0,+\rangle\,.
 \end{equation}
There is one additional operator $\cB_+$ in comparison with \eqref{eq:Vmx} since the dimension of the space $\pi_1^+\cF^+$ is $L+1$ in comparison to $\text{dim}(\pi_1^-\cF^-)=L$.
Then, the computation of the  action of $\cD_+(u,L)$ on this state leads to the computation of $\cD_+(u,0)|0,+\rangle$. Unfortunately this term is ill-defined (see the expression \eqref{eq:defD} with $m=0$).
Therefore, this Bethe state is not a good candidate to be an eigenvalue of $T^+$.

Let us now consider the expression \eqref{eq:T2} between $T^+$ and $\cD_+(u,-L-1)$ and the following Bethe state
\begin{equation}
  W^+(\overline z)= \cB_+(z_0,-L-1)\cB_+(z_1,-L-2) \dots \cB_+(z_{L-2},-2L+1) \cB_+(z_{L-1},-2L)  |0,+\rangle\,,
 \end{equation}
with $\overline z=\{z_0,z_1,\dots ,z_{L-1}\}$.

Proceeding along the lines of the previous paragraph, the eigenvalues of $T^+$ are found to be
\begin{eqnarray}
 t_{\ell, +}(u)&=&2\fc(2u) - 2\Delta(1-u) \prod_{i=0}^{L-1} f(u,z_i)-2\Delta(1+ u) \prod_{i=0}^{L-1} f(- u,z_i)\nonumber\\
 &&-2\frac{\fs(4L+2)\fs(2L+1)\fc(2u(L+1))}{\fc(2L+1)}\prod_{i=0}^{L-1} \frac{1}{4\fs(z_i+u)\fs(z_i-u)} \,. \label{eq:t3}
\end{eqnarray}
with the following Bethe equations
\begin{eqnarray}
  \frac{\Delta(1-z_j)\fs(2z_j-1)}{\Delta(1+z_j)\fs(2z_j+1)}&=&\prod_{\genfrac{}{}{0pt}{}{i=0}{i\neq j}}^{L-1} \frac{\fs(z_j-z_i+1)\fs(z_j+z_i+1)}{\fs(z_j-z_i-1)\fs(z_j+z_i-1)} \label{eq:be3}\\
  &&\hspace{-3cm}-\frac{\fs(4L+2)\fs(2L+1)\fc(2z_j(L+1))\fs(2z_j-1)}{\fc(2L+1)\fs(2z_j+1)\Delta(1+z_j)} \prod_{i=0}^{L-1} \frac{1}{4\fs(z_i-z_j+1)\fs(z_i+z_j-1)}  \,.\nonumber
\end{eqnarray}
For a given Bethe root $\overline{z}$, the function $W^+(\overline{z})$ gives one of the $L$ eigenfunction $|t_{\ell, +}\rangle$ of $T^+$.

\section{Concluding remarks}

Our objective was to demonstrate that the Bethe ansatz methods developed to study integrable systems can be applied in time and band limiting problems. 
We have recalled the definition of these problems in the special case of the discrete Fourier transform and have shown that the associated time and 
band limiting operators could be expressed as polynomials of the commuting matrices. These were recognized as algebraic Heun-Askey-Wilson operators 
and diagonalized by Bethe ansatz. The result was an expression for the eigenvalues of these Heun operators, and by extension for the spectrum of the time and 
band limiting operators, as functions of the solutions of Bethe equations. 

The present work opens the door for further researchs. There exist many different Heun operators and their connection with integrable systems 
remains an interesting open question. Integral operators, with kernel different from the \textit{sinc} one studied in this paper, arise and are studied 
in various domains. It would be interesting to generalize the work done here to these kernels.
There are also numerical and analytical methods developed in both contexts. The comparison between these methods may improve, for example, the resolution of the Bethe equation or 
speed up the reconstruction of a signal from its Fourier data.\\
\vspace{1cm}
 
\noindent
\textbf{Acknowledgments: }
The authors would like to thank warmly F.A. Gr\"unbaum for dicussions.  
PAB holds a scholarship from the Natural Sciences and Engineering
Research Council (NSERC) of Canada. NC is partially supported by Agence
Nationale de la Recherche, Projet ANR-18-CE40-0001. LV gratefully acknowledges a Discovery Grant from NSERC.

\end{document}